\documentclass{ws-ijmpa}
\usepackage[super,compress]{cite}
\usepackage{graphicx}
\usepackage{url}
\usepackage{hyperref}

\begin{document}


\catchline{}{}{}{}{}

\title{Associated production of $J/\psi$ and direct photon in the
 NRQCD and the ICEM using the high-energy factorization}

\author{Lev Alimov}

\address{Samara National Research University, Moskovskoe Shosse,
34, 443086, Samara, Russia.\\
alimov.le@yandex.ru}

\author{Anton Karpishkov}

\address{Samara National Research University, Moskovskoe Shosse,
34, 443086, Samara, Russia }
\address{Joint Institute for Nuclear Research, Dubna, 141980
Russia.\\
karpishkoff@gmail.com}

\author{Vladimir Saleev}

\address{Samara National Research University, Moskovskoe Shosse,
34, 443086, Samara, Russia }
\address{Joint Institute for Nuclear Research, Dubna, 141980
Russia.\\
saleev.vladimir@gmail.com}

\maketitle


\begin{abstract}
We study the associated $J/\psi$ and direct photon production in the
high-energy factorization, as it is formulated in the parton
Reggeization approach, using two different models for the
hadronization of a heavy quark-antiquark pair into a heavy
quarkonium, namely the non-relativistic quantum chromodynamics
(NRQCD) and the improved color evaporation model (ICEM).
We find
essential differences in the predictions for cross-section and
transverse momenta spectra obtained using the NRQCD and the ICEM,
which can be used to discriminate between these models.
Our prediction for cross-sections of the associated $J/\psi$ and direct
photon production at the LHC energies slightly overestimates the
results  obtained early in the next-to-leading order (NLO)
calculation in the collinear parton model (CPM).
We predict
different two-particle correlation spectra in the associated
$J/\psi$ and direct photon production which may be of interest for
experimental study.
\end{abstract}
\maketitle
%

\section{Introduction}
The experimental study of the associated production of $J/\psi$-meson
and direct photon production in high energy proton-proton collisions is of
considerable interest, not only for verifying the predictions of the
perturbative quantum chromodynamics (QCD) and various models of
heavy quark hadronization  into heavy quarkonium
\cite{drees1992associate,mehen1997testing}, but also for obtaining
information about the gluon parton distribution function (PDF) of a
proton, including the transverse momentum dependent (TMD) gluon PDFs
\cite{doncheski1994associated, den2014accessing}.

The value of the strong coupling constant at the scale of a charm
quark mass $\alpha_S(m_c) \simeq 0.3$ is small enough to use
perturbative QCD calculations to describe the charmonium production.
At present, an accuracy level corresponding to next-to-leading order
(NLO) in $\alpha_S$ calculation based on the collinear parton model
(CPM) has been achieved for the prompt $J/\psi$ production
\cite{butenschoen2013next} as well as for associated $J/\psi$ and
direct photon production \cite{li2009next}.

The process of hadronization of a $c\bar c-$pair into the
charmonium state has a non-perturbative nature that can be described only
within the framework of phenomenological models. In the color
singlet model (CSM) \cite{baier1983hadronic,berger1981inelastic}, it
is assumed that a quark-antiquark pair forms a color singlet state
with quantum numbers of the final charmonium. In a more general
framework of the non-relativistic quantum chromodynamics (NRQCD),
which takes into account relativistic corrections in terms of the
relative velocities of the quark and anti-quark inside the charmonium,
the production of charmonium may be via  color-octet intermediate
states \cite{bodwin1995rigorous}. Another approach to describe the
hadronization is the color evaporation model (CEM). It assumes that
all quark-antiquark pairs with an invariant mass between the
thresholds for the charmonium production and the open charm
production convert into bound charmonium states $\cal C$, with a
certain $F^{\cal C}$ conversion probability
\cite{fritzsch1977producing,halzen1977cvc}. Nowadays,  the improved
CEM (ICEM) has been suggested by Ma and Vogt in Ref.
\cite{ICEM2016}.

The crown stone in describing the charmonium production in high
energy proton-proton collisions is the factorization of a hard and a
soft physics phenomena. At the high transverse momentum,
$(p_T>>m_C)$, where the initial parton transverse momenta may be
neglected, the charmonium production in the proton-proton hard
collisions can be described adequately within the CPM
\cite{collins2011foundations}. However, to describe $p_T$-spectrum
at low the transverse momentum ($p_T << m_C$), the approach must be
depended on a non-perturbative transverse momenta of the initial
partons. This is achieved by using the transverse momentum dependent
(TMD) factorization approach, well-known as the TMD parton model, which
takes the effects of intrinsic parton motion into account
\cite{collins1989factorization}. To describe the experimental data
in the intermediate range of transverse momenta, $p_T\simeq m_C$,
different methods are used, combining the results of calculations
based on the CPM and TMD PM \cite{Echevarria:2018qyi}. At high
energies, an alternative method for describing the cross-section
across in the entire range of $p_T$ is using the parton Reggeization
approach (PRA)
\cite{nefedov2013dijet,karpishkov2017angular,nefedov2020high}. The
PRA is a version of the high-energy factorization (HEF) formalism,
which is based on the modified multi-Regge kinematics (MRK)
approximation of the QCD when we deal with the Reggeization of
parton amplitudes. In the PRA, we have described early the
experimental data for prompt and direct $J/\psi$ production at the
energies $\sqrt{s}=1.8-13$ TeV using the NRQCD approach
\cite{Kniehl:2006sk,NSS2012,kniehl2016psi} as well as using the ICEM
\cite{chernyshev2022single}.

At present, a significant amount of experimental data has been
collected for the production of prompt $J/\psi$-mesons in hadronic
interactions at energies ranging from $\sqrt{s}=19$ GeV to
$\sqrt{s}=13$ TeV \cite{Brambilla:2010cs}. The production of a
single direct photons in hadron-hadron collisions has been
extensively studied experimentally over a wide range of energies at
the fixed target experiments~\cite{vogelsang1997compilation} and at
the RICH, Tevatron and LHC colliders
\cite{ATLAS:2021mbt,ALICE:2019rtd,CMS:2018qao}. In the PRA, we
studied single, double, and triple isolated photon production at the
LHC in Refs.
\cite{Saleev:2008cd,Kniehl:2011hc,Nefedov:2015ara,Karpishkov:2022ukm}

However, to date, no measurements have been made for the associated
$J/\psi$ and direct photon production cross-sections. In this paper,
we study the associated production of $J/\psi$-meson and direct
photon using the HEF, as it is formulated in the PRA, using two
different models for the heavy quark hadronization into heavy
quarkonium states, namely the NRQCD and the ICEM. We predict the
cross-section and relevant spectra of $J/\psi$ and direct photon
pairs in proton-proton collisions at the energy $\sqrt{s}=13$ TeV.

\section{Parton Reggeization Approach}
\label{sec:PRA} The PRA is a gauge-invariant version of the
$k_T-$factorization approach, which has been proven in the leading
logarithmic approximation (LLA) of the high energy QCD
\cite{collins1991heavy,catani1994high,gribov1983semihard}. The crown
stones of the PRA are amplitude factorization in the Regge  limit of
the QCD (amplitude Reggeization), the Effective Field Theory (EFT)
of Reggeized gluons and Reggeized quarks, proposed by Lev Lipatov in
Ref. \cite{lipatov1995gauge}, and the modified
Kimber-Martin-Ryskin-Watt (KMRW)
\cite{kimber2001unintegrated,watt2003unintegrated} model for
unintegrated  parton distribution functions (uPDF)
\cite{nefedov2020high}.

In the PRA, the cross-section of the process direct $J/\psi$
production together with large-$p_T$ photon, $pp\to J/\psi\gamma X$,
is expressed as a convolution of the relevant Reggeized
parton-parton cross-section and uPDFs. For example, for the
gluon-gluon fusion subprocess we can write:
\begin{eqnarray}
\label{factorization} d\sigma(pp\to J/\psi\gamma X) &=&
\int\frac{dx_1}{x_1}\int \frac{d^2q_{T1}}{\pi}\Phi_g(x_1,t_1,\mu^2)
\int\frac{dx_2}{x_2}\int
\frac{d^2q_{T2}}{\pi}\Phi_g(x_2,t_2,\mu^2)\times \nonumber \\
& \times & d\hat{\sigma}^{PRA}(RR\to J/\psi\gamma)
\end{eqnarray}
where $q_{1,2}=x_{1,2}P_{1,2}+q_{1,2T}$ are 4-momenta of Reggeized
gluons, $P_{1,2}^\mu=\frac{\sqrt{s}}{2}(1,0,0,\pm 1)$ are 4-momenta
of protons, $q_{1,2T}=(0,{\bf q}_{1,2T},0)$ are transverse 4-momenta
of gluons , $t_{1,2}=-{\bf q}^2_{1,2T}$, $\Phi_{g}(x,t,\mu^2)$ is
the uPDF of the Reggeized gluon. Note, the modified KMRW uPDF
satisfied exact normalization condition at the arbitrary
$x$~\cite{nefedov2020high},
\begin{equation}\int_0^{\mu^2} \Phi_g(x,t,\mu^2) dt=
x f_g(x,\mu^2).
\end{equation}
The patron-parton cross-section $d\hat{\sigma}^{PRA}(RR\to
J/\psi\gamma)$, as well as $d\hat{\sigma}^{PRA}(RR\to \psi'\gamma)$
and $d\hat{\sigma}^{PRA}(RR\to \chi_{cJ}\gamma)$, are written via
the squared Reggeized amplitude $\overline{|M|^2}_{PRA}$ by a usual
textbook formula, see (\ref{commonWay}) and (\ref{commonWay2to3}).

The parton-parton scattering amplitudes in the PRA are calculated
using the Feynman rules of the Lipatov EFT. Within this framework,
the amplitudes are gauge-invariant and the initial-state partons are
treated as Reggeized partons. To obtain  Reggeized amplitudes, we
use in our calculations the {\texttt Mathematica} package {\texttt
FeynArts}~\cite{hahn2001generating} and the corresponding model file
{\texttt ReggeQCD} by Maxim Nefedov~\cite{karpishkov2017angular}.

The gauge invariance of all amplitudes has been verified
additionally. Furthermore, the squared amplitudes in the PRA have an
explicit collinear limit, which has been verified analytically for
each considered squared amplitudes as follows
\begin{equation}
\label{collinearLimitForMatrix} \lim\limits_{t_1,t_2\to
0}\int\limits_{0}^{2\pi}\int\limits_{0}^{2\pi}\frac{d\phi_1d\phi_2}{(2\pi)^2}
\overline{|M|^2}_{PRA}=\overline{|M|^2}_{CPM}.
\end{equation}

The PRA has been used to describe the production of direct and
prompt $\mathrm{J/\psi}$ mesons in high energy proton-proton
collisions. In previous studies, we have found good agreement
between the LO PRA computation and experimental data from the CDF,
ATLAS, CMS, and LHCb collaborations
\cite{Kniehl:2006sk,PhysRevD.85.074013,NSS2012,
nefedov2013charmonium,kniehl2016psi,Saleev:2012hi}.


\section{PRA using NRQCD}

The approach of NRQCD is a theoretical framework that separates the
effects of short-distance and long-distance physics. The cross
section of the charmonium $\cal C$ production in the gluon-gluon
fusion can be expressed as a sum over all possible states of a
$c\bar{c}$-pair with respect to the quantum numbers, see
Ref.~\cite{bodwin1995rigorous}:
\begin{equation}
\sigma(RR\to {\cal C}\gamma)=\sum\limits_n \hat{\sigma}(RR\to c
\bar{c}[n]\gamma) \frac{\langle {\large O}^{\cal
C}[n]\rangle}{N_{col}N_{pol}}
\end{equation}
where  $[n] = [{}^{2S+1}L_J^{(1,8)}]$ is the state of a
$c\bar{c}$-pair, written using the usual spectroscopic notation. The
quantum number in upper index  $^{(1,8)}$ identifies color singlet
or color octet states, respectively. The $\hat{\sigma}$ represents
the partonic cross-section for the production of the state $c\bar{c}
[n]$, and ${\langle O}^{\cal C}[n] \rangle$ are the long-distance
matrix element (LDME), which describes the transition of an
intermediate state into a charmonium $\cal C$. One has
$N_{col}=2N_c$ for the color-singlet states and $N_{col}=N_c^2-1$
for the color-octet states, $N_{pol}=2J+1$.

To study prompt $J/\psi$ production we take into consideration the
direct production in the subprocess
\begin{equation}
R+R \to J/\psi + \gamma \label{eq:direct},
\end{equation}
and the cascade production in the subprocesses
\begin{eqnarray}
R+R &\to & \psi' + \gamma \label{eq:psi2}\\
R+R & \to & \chi_{cJ} + \gamma, \label{eq:chic}
\end{eqnarray}
via $\psi'\to J/\psi X$ and $\chi_{cJ}\to J/\psi \gamma$ decays.  At
the LHC energies $J/\psi+\gamma$ production subprocesses in the
quark-antiquark annihilation give small contributions and may be
neglected. We derive analytical formulae for the relevant Reggeized
amplitudes using {\texttt FeynArts} and {\texttt ReggeQCD} packages.
The corresponding squared amplitudes are unwieldy for the
presentation here and they may be obtained from authors by the
request.

The master formula for the numerical calculation in the PRA can be
obtained from the factorization formula (\ref{factorization}) and
the partonic cross-section
\begin{equation}
\label{commonWay} d\hat\sigma^{PRA}(RR\to
J/\psi\gamma)=(2\pi)^4\delta^{(4)}(q_1+q_2-p_\psi-k_\gamma)\frac{\overline{|M|^2}_{PRA}}{I}\frac{d^3p_\psi}{(2\pi)^3
2p^0_{\psi}}\frac{d^3k_{\gamma}}{(2\pi)^3 2k^0_{\gamma }},
\end{equation}
where $I= 2x_1x_2s$ is the flux factor,  $p^\mu_\psi$ is the
$J/\psi$ 4-momentum, $k_\gamma$ is the photon 4-momentum. In such a
way, in the PRA using the NRQCD we can write for direct $J/\psi +
\gamma$ production cross-section
\begin{equation}
\frac{d\sigma(pp\to J/\psi \gamma X)}{dp_{\psi T}dy_{\psi}dk_{\gamma
T}dy_{\gamma}d\Delta\phi}=\frac{p_{\psi T} k_{\gamma T}}{16\pi^3}
\int dt_1 \int d\phi_1 \Phi_g(x_1,t_1,\mu^2)\Phi_g(x_2,t_2,\mu^2)
\frac{\overline{|M|^2}_{PRA}}{(x_1 x_2 s)^2}
\end{equation}
where ${\bf q}_{2T}={\bf p}_T+{\bf k}_T-{\bf q}_{1T}$,
$x_1=(p^0_{\psi }+k^0_{\gamma }+p^z_{\psi }+k^z_{\gamma
})/\sqrt{s}$, $x_2=(p^0_{\psi }+k^0_{\gamma }-p^z_{\psi
}-k^z_{\gamma })/\sqrt{s}$, $y_\psi$ is the $J/\psi$ rapidity,
$y_\gamma$ is the photon rapidity,
$\Delta\phi=\phi_\psi-\phi_\gamma$. Squared amplitudes
$\overline{|M|^2}_{PRA}$ are written as functions of conventional
Mandelstam variables $\hat s, \hat t, \hat u$ and variables
$t_1,t_2,a_k,a_p,b_k,b_p$, where $a_k=2(k_\gamma.P_2)/s,
a_p=2(p_\psi.P_2)/s, b_k=2(k_\gamma.P_1)/s, b_p=2(p_\psi.P_1)/s$.

\section{PRA using ICEM}

In the PRA using the ICEM, the description of the associated
production of $J/\psi$ and large-$p_T$ direct photon at the leading
order in $\alpha_S$ is possible via the subprocesses
\begin{equation}R + R
\to c+\bar{c}+ \gamma \end{equation}\label{eq:ccg} and
\begin{equation}Q + \bar Q \to
c+\bar{c}+ \gamma. \label{qqg} \end{equation}
 However, the latter subprocess
contributes negligibly (see Fig. \ref{ris:14tev_octet}), and
therefore, it can be ignored.

In the ICEM, the  prompt $J/\psi$-meson production cross-section may
be written in the following manner:
\begin{equation}
    \label{ICEM}
    \sigma(pp\to J/\psi\gamma X)=F^{\psi}
    \int\limits_{m_\psi^2}^{4 m_D^2 }\frac{d\sigma(pp\to c\bar{c}\gamma
    X)}{dM^2}dM^2
\end{equation}
where $M$ is the $c\bar{c}$-pair invariant mass, $m_D$ is the mass
of the lightest $D$-meson. In other words, the integration is
carried out from the charmonium mass up to the open charm production
threshold. The ICEM takes into account the fact that the mass of the
intermediate state (i.e. the invariant mass of the $c\bar{c}$-pair)
differs from the mass of the $J/\psi$ meson and the relation between
4-momenta should be accounted $p_\psi^\mu=p^\mu \frac{m_{\psi}}{M}$,
here $p^\mu=p_{c}^\mu+p_{\bar{c}}^\mu$. In the study of prompt
$J/\psi$ production at the LHC using the PRA plus ICEM approach, it
was found that a good description of the data may be archived with
hadronization parameter $F^{\psi} \simeq 0.02$ at the $\sqrt{s}=13$
TeV~\cite{chernyshev2022single}.

The cross-section for the subprocess (\ref{eq:ccg}) is written  the
same as (\ref{commonWay}) taken into account that it is $2\to 3$
parton-level process:
\begin{eqnarray}
    \label{commonWay2to3}
     d\hat{\sigma}^{PRA}(R+R\to c+\bar{c}+\gamma)&=&(2\pi)^4\delta^{(4)}(q_1+q_2-p_c-p_{\bar{c}}-k_\gamma)
     \frac{\overline{|M|^2}_{PRA}}{2x_1x_2s} \times \nonumber \\
    & \times & \frac{d^3p_{c}}{(2\pi)^3 2p_{c0}}\frac{d^3p_{\bar{c}}}{(2\pi)^3 2p_{\bar{c}0}}\frac{d^3k_\gamma}{(2\pi)^3
     2k_{\gamma 0}}.
\end{eqnarray}

The master formula for numerical calculations in the PRA using the
ICEM, obtained from (\ref{ICEM}) and (\ref{commonWay2to3}), reads
\begin{eqnarray}
\frac{d\sigma(pp\to J/\psi\gamma X)}{dp_{\psi T}dy_{\psi}dk_{\gamma
T}dy_{\gamma}d\Delta\phi}&=& F^{\psi}\times \frac{p_{\psi
 T} k_{\gamma T}}{1024\pi^6} \int^{4m_D^2}_{m_\psi} dM^2 \int dt_1 \int d\phi_1 \int
d\Omega_{c\bar c}  \left(\frac{M}{m_\psi}\right)^2 \times \nonumber
\label{eq:PRAICEM}
\\
&& \times \sqrt{1-\frac{4m_c^2}{M^2}}
\Phi_g(x_1,t_1,\mu^2)\Phi_g(x_2,t_2,\mu^2)
\frac{\overline{|M|^2}_{PRA}}{(x_1 x_2 s)^2},
\end{eqnarray}
where $d\Omega_{c\bar c}=\sin (\theta)d\theta d\phi$, angles
$\theta$ and $\phi$ are associated with the rest frame of the
$c\bar{c}$-pair~\cite{nefedov2020high}, and we put during numerical
calculations the mass of $c-$quark is equal $m_c=1.3$ GeV, the mass
of $D-$meson - $m_D=1.86$ GeV, and the mass of $J/\psi$ -
$m_\psi=3.097$ GeV.

It is suitable for numerical calculation to rewrite $c-$quark
(antiquark) 4-momenta as follows,
\begin{equation} p_c^\mu=\frac{1}{2}p^\mu+r^\mu \mbox{ and } p_{\bar
c}^\mu=\frac{1}{2}p^\mu-r^\mu ,\end{equation}
 where 4-momentum of
relative motion $r^\mu$ is written in terms of invariant mass $M$,
$c\bar c-$pair transverse momentum $p_T$ and $c\bar c-$pair rapidity
$y$. Such a way,
\begin{equation}
r^\mu = \frac{1}{2}\sqrt{M^2-4 m_c^2} \left( X^\mu
\sin(\theta)\cos(\phi)+ Y^\mu \sin(\theta)\sin(\phi)+Z^\mu
\cos(\theta)\right),
\end{equation}
where
\begin{eqnarray}
X^\mu &=&\frac{1}{M} \left( p_{T} \cosh(y), \sqrt{M^2+p_{T}^2},0,p_{T} \sinh(y)\right), \nonumber\\
Y^\mu &=& \mbox{sign}(y) \left(0,0,1,0\right), \nonumber \\
Z^\mu &=& \mbox{sign}(y) \left(\sinh(y),0,0,\cosh(y)\right).
\nonumber
\end{eqnarray}

The squared amplitude $ \overline{|M|^2}_{PRA}$ is calculated using
{\texttt FeynArts} and {\texttt ReggeQCD} packages and may be
written as a function of $\hat s=(q_1+q_2)^2$, $\hat t=(q_1-k)^2$,
$\hat u=(q_2-k)^2$, $w_1=(q_1-p_c)^2$, $w_2=(q_2-p_{\bar c})^2$,
$a_c=2(p_c.P_2)/s$, $a_{\bar c}=2(p_{\bar c}.P_2)$,
$a_k=2(k_\gamma.P_2)$, $b_c=2(p_c.P_1)/s$, $b_{\bar c}=2(p_{\bar
c}.P_1)$, $b_k=2(k_\gamma.P_1)$.

The calculation in the PRA using the ICEM, as they are explained
above, may be performed in a different way within the Monte-Carlo
event generator {\texttt KaTie}~\cite{katie}, the same as in
Ref.~\cite{Chernyshev:2023kzk} for the process of the associated
$J/\psi$ plus $Z(W)$-boson production. We have done cross-check all
calculations in the PRA using the ICEM within the {\texttt KaTie},
and found a good agreement.

\section{Results}

Now, we are in a position to discuss the results of the numerical
calculations in which we compare predictions obtained in the PRA
using the NRQCD and in the PRA using the ICEM.

\begin{figure}[h]
\begin{center}
     \includegraphics[width=1.0\linewidth]{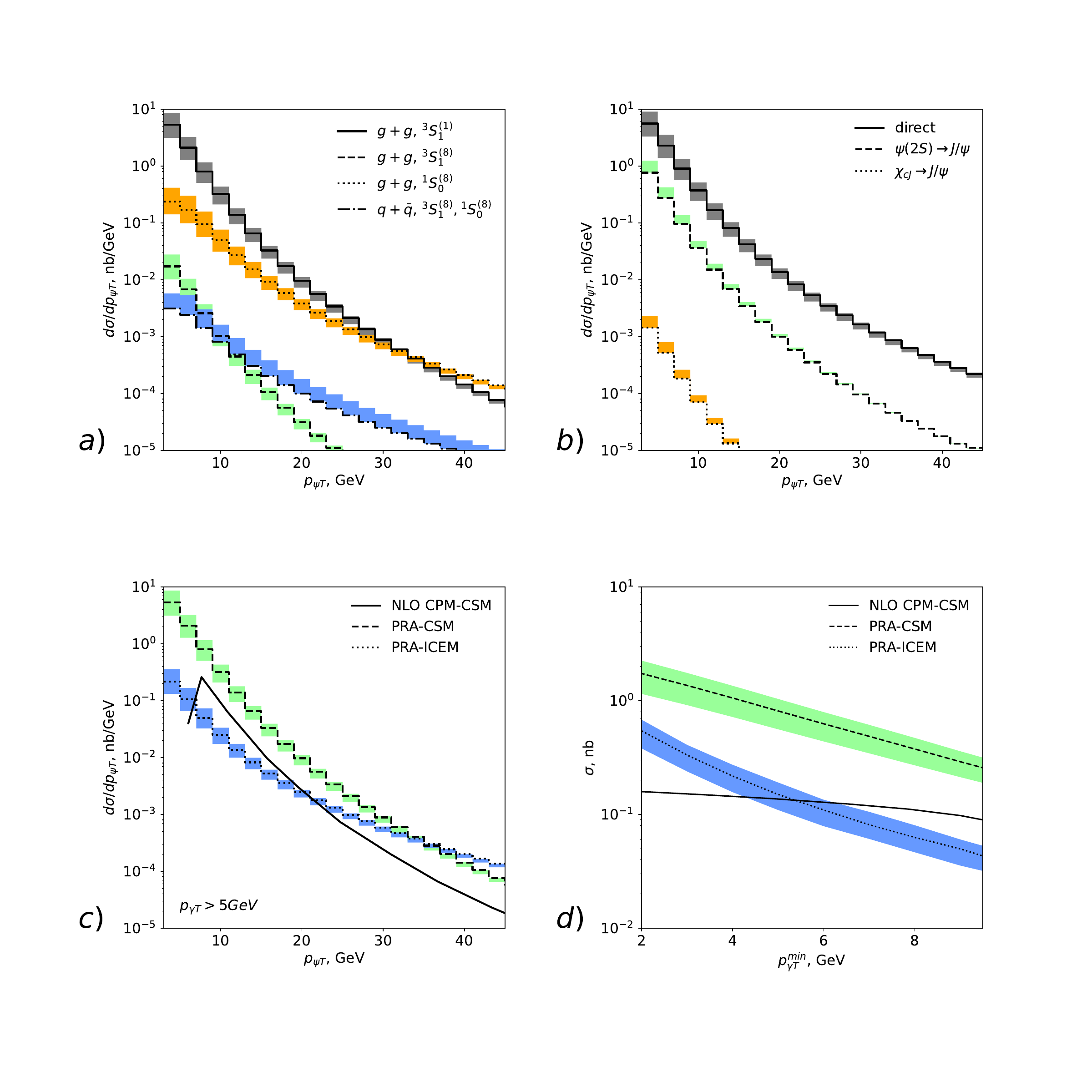}
\caption{Cross-section for the associated production of
$J/\psi+\gamma$ as a function of the $p_{\psi T}$ at the
$\sqrt{s}=14$ TeV obtained in the PRA using the NRQCD at
$|y_{\gamma, \psi}|<2$. In the panel $(a)$, solid curve --- the CSM
contribution, dashed curve
--- the contribution of the color octet $[^1S_0^{(8)}]$
state, dotted curve --- the contribution of color-octet
$[^3S_1^{(8)}]$ state and  the contribution of quark-antiquark
annihilation processes are shown as a dashed blue curve. In the panel
(b), solid curve
--- the direct $J/\psi$ production, dashed curve --- the cascade via
$\psi(2S)\to J/\psi X$ production, dotted curve --- the cascade via
$\chi_{cJ}\to J/\psi \gamma$ production. In the panel (c), solid
curve --- the NLO CPM using the CSM calculation from
Ref.~\cite{li2009next}, dashed curve
--- the PRA using the CSM calculation and dotted curve --- the PRA using the ICEM calculation.
In the panel (d), cross-section for the associated production of
$J/\psi+\gamma$ as a function of the $p_{\gamma T min}$ at the
$p_{\psi T}>10$ GeV. Solid curve --- the NLO CPM using the CSM
calculation from Ref.~\cite{li2009next}, dashed curve --- the PRA
using the CSM calculation and dotted curve --- the PRA using the
ICEM calculation. }
        \label{ris:14tev_octet}
\end{center}
\end{figure}

In the first step, we calculate $J/\psi+\gamma$ production cross
section in the PRA using the NRQCD when $J/\psi$ is produced directly
via different intermediate color-singlet and color-octet states. The
results are shown in Fig.~\ref{ris:14tev_octet}(a). We confirm the
well-known conclusion on the dominant role of color-singlet production
mechanism in the direct $J/\psi+\gamma$ production~
\cite{doncheski1994associated,drees1992associate}. In
Fig.~\ref{ris:14tev_octet} (a), we plot total
contribution from the quark-antiquark annihilation subprocesses
included both color-singlet and color-octet production mechanisms additionally
and find that it is very small at the $\sqrt{s}=13-14$ TeV.
 Taking in mind
experimental difficulties in the separation of the direct and the prompt
$J/\psi$ production processes, we estimate $J/\psi+\gamma$
production cross-section when $J/\psi$ is produced in cascade
processes via decays of the exited charmonium state, $\psi(2S)\to
J/\psi X$ and $\chi_{cJ}\to J/\psi \gamma$. We see in Fig.
\ref{ris:14tev_octet} (b) that only $J/\psi$ production via
$\psi(2S)$ decay may be important with the contribution in a few
percent in the prompt $J/\psi+\gamma$ production cross-section. Such
a way, we will take into account the NRQCD approach only
color-singlet contribution, i.e. we will use the
CSM~\cite{berger1981inelastic,baier1983hadronic}. The contributions
from cascade productions will be also neglected such as small ones
in comparing with the theoretical uncertainties of the PRA calculations.

To validate our results additionally, we compare them with the NLO
CPM using the CSM calculations at the energy $\sqrt{s}=14$ TeV
published in Ref. \cite{li2009next}. We calculate the transverse
momentum spectrum of the $J/\psi$-mesons for the kinematic
conditions $|y_\psi|,|y_\gamma| < 3$ and $p_{T\gamma}
>5$ GeV, which were used in Ref.~\cite{li2009next}. We find the PRA
using the CSM calculation sufficiently overestimates the NLO CPM
using the CSM cross-section at all $J/\psi$ transverse momenta, see
Fig.~\ref{ris:14tev_octet} (c). This is an interesting finding because
in the single $J/\psi$ production the results obtained in the LO PRA
and the NLO CPM using the NRQCD are approximately coincided. The
significant disparity has also been identified in the PRA
predictions made using different hadronization models, the
$p_T-$spectrum obtained in the ICEM is sufficiently lower than the
spectrum obtained in the CSM starting from small $p_{\psi T}$ up to
$p_{\psi T}=30$ GeV. The NLO CPM using the ICEM calculations are
absent and we can't compare NLO CPM predictions obtained using
different hadronization models.

The total cross-sections as functions of the $p_{T\gamma}^{min}$ are
shown in Fig.~\ref{ris:14tev_octet}(d). The prediction of the PRA
using the CSM is larger the prediction of the NLO CPM using the CSM
at small $p_{T\gamma}^{min}$ because we integrate over the
$p_{T\psi} > 10$ GeV where the NLO CPM using the CSM prediction
strongly suppressed when $p_{T\gamma}$ is small due to the mostly
back-to-back production of the $J/\psi$ and photon is dominant in the
CPM instead of the PRA. In Fig.~\ref{ris:14tev_octet} and below, the
theoretical uncertainties due to variations in the hard scale by a
factor of 2 is indicated by shaded regions.

\begin{figure}[h]
\begin{center}
\includegraphics[width=0.9\linewidth]{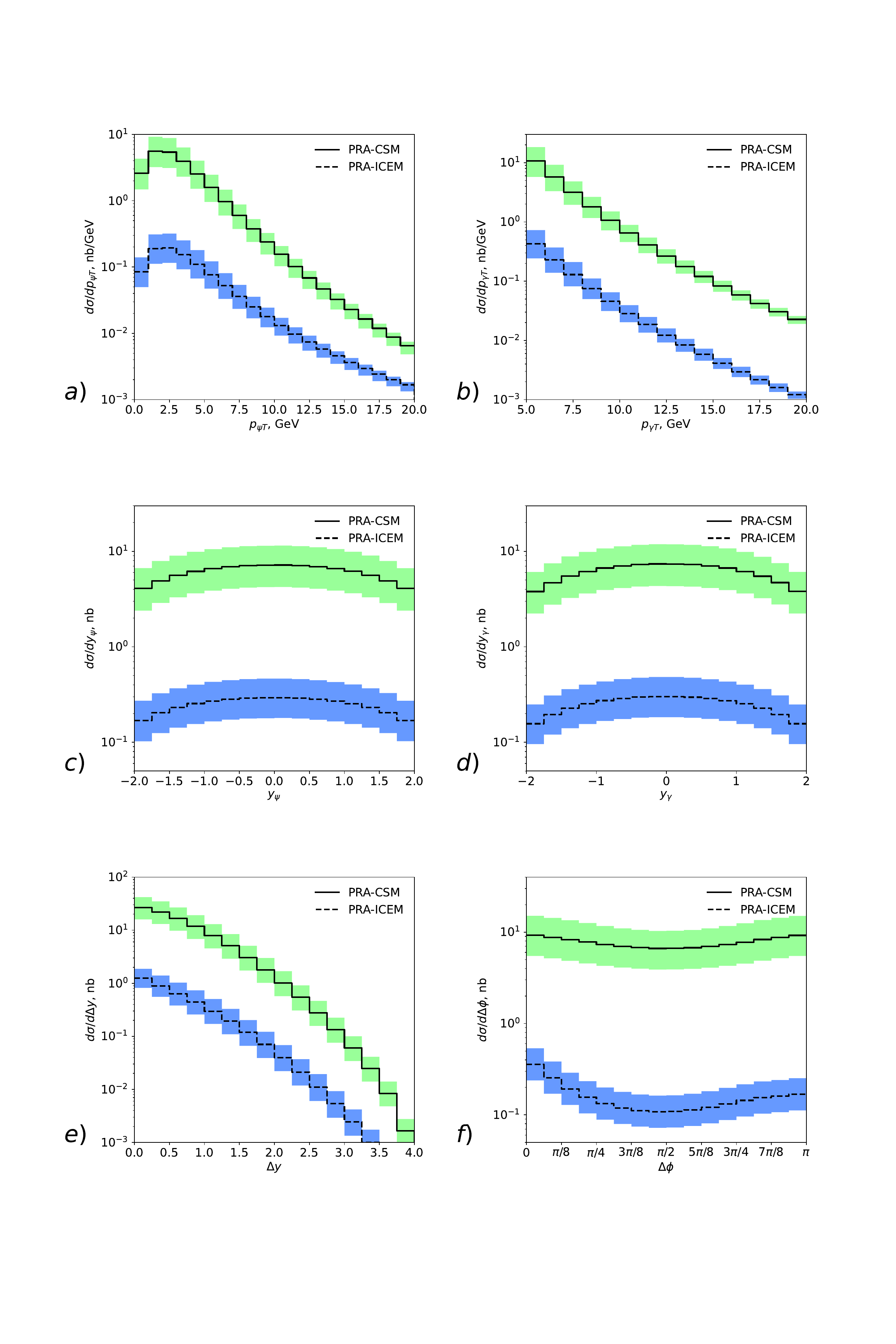}
\caption{Differential cross-sections for the associated production
of $J/\psi+\gamma$ at the $\sqrt{s}=13$ TeV and $|y_{\gamma,
\psi}|<2$  as functions of the  $p_{\psi T}$ $(a)$,  $p_{\gamma T}$
$(b)$,  $y_\psi$ $(c)$,  $y_\gamma$ $(d)$,  rapidity difference
$\Delta y$ $(e)$ and  azimuthal angle difference $\Delta \phi$
$(f)$. Solid curves are the PRA using CSM calculations, dashed
curves are the PRA using ICEM calculations.}
        \label{ris:LHC13_1}
\end{center}
\end{figure}

\begin{figure}[h]
\begin{center}
\includegraphics[width=1\linewidth]{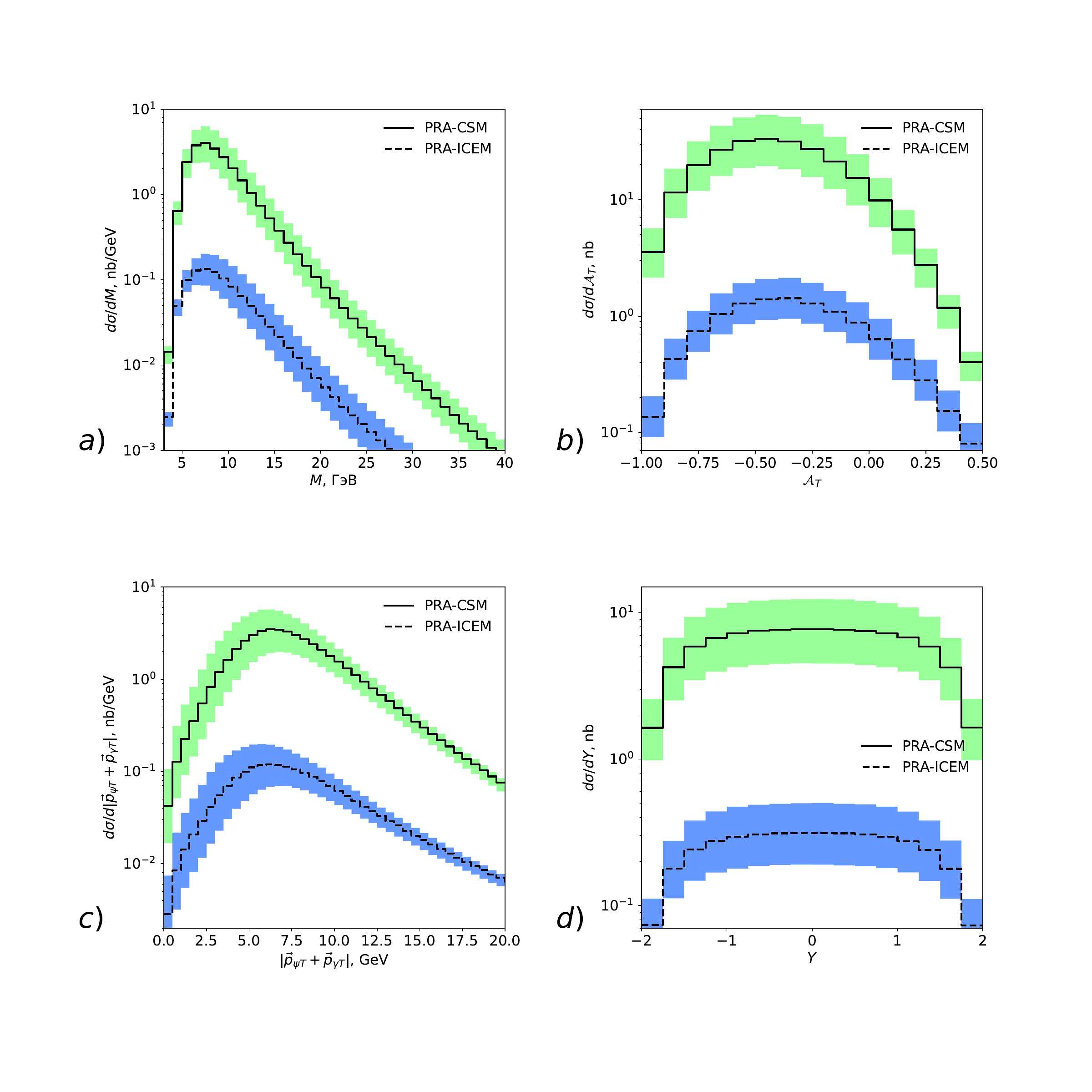}
\caption{Differential cross-sections for the associated production
of $J/\psi+\gamma$ at the $\sqrt{s}=13$ TeV and $|y_{\gamma,
\psi}|<2$  as functions of the  invariant mass $M=M_{\psi\gamma}$
$(a)$,  transverse momentum difference ${\cal A}_T$ $(b)$, pair
transverse momentum $p_T=|{\bf p}_{\psi T}+{\bf p}_{\gamma T}|$
$(c)$ and  pair rapidity $Y_{\gamma\psi}$ $(d)$. Solid curves are
the PRA using CSM calculations, dashed curves are the PRA using ICEM
calculations.}
        \label{ris:LHC13_2}
\end{center}
\end{figure}

Taking into account the relatively small contribution of color-octet
states and cascade production of $J/\psi$-meson, the predictions
were made in the PRA using the CSM for the LHC energy $\sqrt{s} =
13$ TeV in the central rapidity region $|y_{J/\psi}|$ and
$|y_\gamma| < 2$. In Fig.~\ref{ris:LHC13_1}, the differential cross
sections are shown as functions of the $J/\psi$ transverse momentum
$p_{\psi T}$, photon transverse momentum  $p_{\gamma T}$, $J/\psi$
rapidity $y_\psi$, photon rapidity $y_\gamma$, rapidity difference
$\Delta y=|y_\psi-y_\gamma|$ and azimuthal angle difference $\Delta
\phi=|\phi_\psi- \phi_{\gamma}|$. Solid curves are the PRA using CSM
calculations, dashed curves are the PRA using ICEM calculations.
 In Fig.~\ref{ris:LHC13_2}, the differential cross-sections are
shown as functions of the invariant mass
$M=M_{\psi\gamma}$, transverse momentum difference ${\cal
A}_T=(|{\bf p}_{\psi T}|+|{\bf p}_{\gamma T}|)/(|{\bf p}_{\psi
T}|+|{\bf p}_{\gamma T}|)$, pair transverse momentum $p_T=|{\bf
p}_{\psi T}+{\bf p}_{\gamma T}|$ and pair rapidity
$Y=Y_{\gamma\psi}$.

\section{Conclusions}

Working in the PRA, we confirm previously obtained results that in
the process of the associated $J/\psi$ and direct photon production
the CSM approximation of the NRQCD is the dominant contribution and we
can neglect color-octet contributions. The second general finding is
that we can neglect small contributions from the quark-antiquark
annihilation processes to calculate $J/\psi$ plus photon production
cross-section at the energy $\sqrt{s}=13-14$ TeV. We find surprising
sufficient differences in predictions based on the CSM and the ICEM
which become larger when the photon transverse momentum increases.
The ICEM prediction is strongly suppressed relative the CSM
prediction instead of good agreement between the ICEM and the NRQCD
calculations for the single $J/\psi$ prompt production. In such a way,
experimental measurement of the $J/\psi$ plus large-$p_T$ photon
production cross-section could be potentially used to distinguish
between the ICEM and the NRQCD.

\section{Acknowledgments}
 The work is supported by the Foundation for the Advancement of
Theoretical Physics and Mathematics BASIS, grant No. 24--1--1--16--5
and by the grant of the Ministry of Science and Higher Education of
the Russian Federation, No. FSSS--2024--0027.

\newpage
\bibliographystyle{ws-mpla}
\bibliography{psi_gamma_mpla}

\begin{thebibliography}{10}

\bibitem{drees1992associate}
M.~Drees and C.~S. Kim, {\em Z. Phys. C} {\bf 53}, 673  (1992).

\bibitem{mehen1997testing}
T.~Mehen, {\em Phys. Rev. D} {\bf 55}, 4338  (1997).

\bibitem{doncheski1994associated}
M.~A. Doncheski and C.~S. Kim, {\em Phys. Rev. D} {\bf 49}, 4463  (1994).

\bibitem{den2014accessing}
W.~J. den Dunnen, J.~P. Lansberg, C.~Pisano and M.~Schlegel, {\em Phys. Rev.
  Lett.} {\bf 112},   212001  (2014).

\bibitem{butenschoen2013next}
M.~Butenschoen and B.~A. Kniehl, {\em Mod. Phys. Lett. A} {\bf 28},   1350027
  (2013).

\bibitem{li2009next}
R.~Li and J.-X. Wang, {\em Phys. Lett. B} {\bf 672}, 51  (2009).

\bibitem{baier1983hadronic}
R.~Baier and R.~Ruckl, {\em Z. Phys. C} {\bf 19},   251  (1983).

\bibitem{berger1981inelastic}
E.~L. Berger and D.~L. Jones, {\em Phys. Rev. D} {\bf 23}, 1521  (1981).

\bibitem{bodwin1995rigorous}
G.~T. Bodwin, E.~Braaten and G.~P. Lepage, {\em Phys. Rev. D} {\bf 51}, 1125
  (1995), [Erratum: Phys.Rev.D 55, 5853 (1997)].

\bibitem{fritzsch1977producing}
H.~Fritzsch, {\em Phys. Lett. B} {\bf 67}, 217  (1977).

\bibitem{halzen1977cvc}
F.~Halzen, {\em Phys. Lett. B} {\bf 69}, 105  (1977).

\bibitem{ICEM2016}
Y.-Q. Ma and R.~Vogt, {\em Phys. Rev. D} {\bf 94},   114029  (2016).

\bibitem{collins2011foundations}
J.~Collins, {\em {Foundations of Perturbative QCD}}, Cambridge Monographs on
  Particle Physics, Nuclear Physics and Cosmology, Vol.~32 (Cambridge
  University Press, 7 2023).

\bibitem{collins1989factorization}
J.~C. Collins, D.~E. Soper and G.~F. Sterman, {Factorization of Hard Processes
  in QCD} 1989 pp. 1--91.

\bibitem{Echevarria:2018qyi}
M.~G. Echevarria, T.~Kasemets, J.-P. Lansberg, C.~Pisano and A.~Signori, {\em
  Phys. Lett. B} {\bf 781}, 161  (2018).

\bibitem{nefedov2013dijet}
M.~A. Nefedov, V.~A. Saleev and A.~V. Shipilova, {\em Phys. Rev. D} {\bf 87},
  094030  (2013).

\bibitem{karpishkov2017angular}
A.~V. Karpishkov, M.~A. Nefedov and V.~A. Saleev, {\em Phys. Rev. D} {\bf 96},
   096019  (2017).

\bibitem{nefedov2020high}
M.~A. Nefedov and V.~A. Saleev, {\em Phys. Rev. D} {\bf 102},   114018  (2020).

\bibitem{Kniehl:2006sk}
B.~A. Kniehl, D.~V. Vasin and V.~A. Saleev, {\em Phys. Rev. D} {\bf 73},
  074022  (2006).

\bibitem{NSS2012}
V.~A. Saleev, M.~A. Nefedov and A.~V. Shipilova, {\em Phys. Rev. D} {\bf 85},
  074013  (2012).

\bibitem{kniehl2016psi}
B.~A. Kniehl, M.~A. Nefedov and V.~A. Saleev, {\em Phys. Rev. D} {\bf 94},
  054007  (2016).

\bibitem{chernyshev2022single}
A.~A. Chernyshev and V.~A. Saleev, {\em Phys. Rev. D} {\bf 106},   114006
  (2022).

\bibitem{Brambilla:2010cs}
N.~Brambilla {\em et~al.}, {\em Eur. Phys. J. C} {\bf 71},   1534  (2011).

\bibitem{vogelsang1997compilation}
W.~Vogelsang and M.~R. Whalley, {\em J. Phys. G} {\bf 23}, A1  (1997).

\bibitem{ATLAS:2021mbt}
ATLAS Collaboration, G.~Aad {\em et~al.}, {\em JHEP} {\bf 11},   169  (2021).

\bibitem{ALICE:2019rtd}
ALICE Collaboration, S.~Acharya {\em et~al.}, {\em Eur. Phys. J. C} {\bf 79},
  896  (2019).

\bibitem{CMS:2018qao}
CMS Collaboration, A.~M. Sirunyan {\em et~al.}, {\em Eur. Phys. J. C} {\bf 79},
   ~20  (2019).

\bibitem{Saleev:2008cd}
V.~A. Saleev, {\em Phys. Rev. D} {\bf 78},   114031  (2008).

\bibitem{Kniehl:2011hc}
B.~A. Kniehl, V.~A. Saleev, A.~V. Shipilova and E.~V. Yatsenko, {\em Phys. Rev.
  D} {\bf 84},   074017  (2011).

\bibitem{Nefedov:2015ara}
M.~Nefedov and V.~Saleev, {\em Phys. Rev. D} {\bf 92},   094033  (2015).

\bibitem{Karpishkov:2022ukm}
A.~Karpishkov and V.~Saleev, {\em Phys. Rev. D} {\bf 106},   054036  (2022).

\bibitem{collins1991heavy}
J.~C. Collins and R.~K. Ellis, {\em Nucl. Phys. B} {\bf 360}, 3  (1991).

\bibitem{catani1994high}
S.~Catani and F.~Hautmann, {\em Nucl. Phys. B} {\bf 427}, 475  (1994).

\bibitem{gribov1983semihard}
L.~V. Gribov, E.~M. Levin and M.~G. Ryskin, {\em Phys. Rept.} {\bf 100}, 1
  (1983).

\bibitem{lipatov1995gauge}
L.~N. Lipatov, {\em Nucl. Phys. B} {\bf 452}, 369  (1995).

\bibitem{kimber2001unintegrated}
M.~A. Kimber, A.~D. Martin and M.~G. Ryskin, {\em Phys. Rev. D} {\bf 63},
  114027  (2001).

\bibitem{watt2003unintegrated}
G.~Watt, A.~D. Martin and M.~G. Ryskin, {\em Eur. Phys. J. C} {\bf 31}, 73
  (2003).

\bibitem{hahn2001generating}
T.~Hahn, {\em Comput. Phys. Commun.} {\bf 140}, 418  (2001).

\bibitem{PhysRevD.85.074013}
V.~A. Saleev, M.~A. Nefedov and A.~V. Shipilova, {\em Phys. Rev. D} {\bf 85},
  074013  (2012).

\bibitem{nefedov2013charmonium}
M.~A. Nefedov, V.~A. Saleev and A.~V. Shipilova, {\em Phys. Atom. Nucl.} {\bf
  76}, 1546  (2013).

\bibitem{Saleev:2012hi}
V.~A. Saleev, M.~A. Nefedov and A.~V. Shipilova, {\em Phys. Rev. D} {\bf 85},
  074013  (2012).

\bibitem{katie}
A.~van Hameren, {\em Comput. Phys. Commun.} {\bf 224}, 371  (2018).

\bibitem{Chernyshev:2023kzk}
A.~Chernyshev and V.~Saleev, {\em Int. J. Mod. Phys. A} {\bf 38},   2350193
  (2023), \href{http://arxiv.org/abs/2304.07481}{{\ttfamily arXiv:2304.07481
  [hep-ph]}}.

\end{thebibliography}

\newpage

\end{document}